\documentclass[showpacs,preprintnumbers,amsmath,amssymb,floatfix,prd]{revtex4}
\usepackage{graphicx}
\usepackage{dcolumn}
\usepackage{bm}
\newcommand{\be}{\begin{equation}}

\newcommand{\ee}{\end{equation}}
\newcommand{\bea}{\begin{eqnarray}}
\newcommand{\eea}{\end{eqnarray}}

\newcommand{\nn}{\nonumber}
\newcommand{\de}{\partial}
\newcommand{\besplit}{\begin{split}}
\newcommand{\esplit}{\end{split}}

\newcommand{\al}{\alpha}
\newcommand{\m}{\mu}

\newcommand{\D}{\Delta}
\newcommand{\e}{\epsilon}

\newcommand{\vth}{\vartheta}

\begin{document}

\title{Neutrino emission from compact stars and inhomogeneous color superconductivity
}
\date{\today}
\author{Roberto~Anglani}\email{roberto.anglani@ba.infn.it}
\affiliation{Universit\`a di Bari, I-70126 Bari, Italia}
\affiliation{I.N.F.N., Sezione di Bari, I-70126 Bari, Italia}
\author{Massimo~Mannarelli}
\email{massimo@lns.mit.edu} \affiliation{Center for Theoretical
Physics, Massachusetts Institute of Technology, Cambridge, MA 02139}
\author{Giuseppe~Nardulli}
\email{giuseppe.nardulli@ba.infn.it}\affiliation{Universit\`a di
Bari, I-70126 Bari, Italia} \affiliation{I.N.F.N., Sezione di Bari,
I-70126 Bari, Italia}\author{Marco~Ruggieri}
\email{marco.ruggieri@ba.infn.it}  \affiliation{Universit\`a di
Bari, I-70126 Bari, Italia} \affiliation{I.N.F.N., Sezione di Bari,
I-70126 Bari, Italia}

\preprint{BARI-TH/06-539} \preprint{MIT-CTP 3758}

\begin{abstract}
 We discuss specific heat and neutrino emissivity due to direct Urca
processes for quark matter in the  color superconductive
Larkin-Ovchinnikov-Fulde-Ferrell (LOFF) phase of
Quantum-Chromodynamics. We assume that the three light quarks
$u,\,d,\,s$ are in a color and electrically neutral state and
interact by a four fermion Nambu-Jona Lasinio coupling. We study a
LOFF state characterized by a single plane wave for each pairing.
From the evaluation of neutrino emissivity and fermionic specific
heat, the cooling rate of simplified models of compact stars with a
quark core in the  LOFF state is estimated.
\end{abstract}

 \maketitle
\section{Introduction}
Neutrino emission due to direct Urca processes, when kinematically
allowed, is the most efficient  cooling mechanism for a neutron star
in the early stage of its lifetime \cite{ShapiroTeukolsky}. After a
very short epoch, when the temperature of the compact star is of the
order of $\sim 10^{11}$K and neutrinos are trapped in the stellar
core~\cite{Ruster:2005ib}, the temperature drops and neutrinos are
able to escape. However, even for smaller temperatures, e.g. below
$10^9$K, the direct nuclear Urca processes $n \to p + e + \bar\nu_e$
and $e^- +p \to n + \nu_e$, which would produce rapid cooling, are
not kinematically allowed, because energy and momentum cannot be
simultaneously conserved. Therefore only modified Urca processes can
take place, where a bystander particle, present in the reaction,
allows energy-momentum conservation. The resulting cooling is less
rapid because neutrino emission rates turn out to be
$\varepsilon_{\nu} \sim T^8$, much smaller than the emission rate
$\varepsilon_{\nu} \sim T^6$ due to direct Urca processes.

These considerations apply to stars containing only nuclear
matter. If hadronic densities in the core of neutron stars are
sufficiently large, the central region of the star should consist
of deconfined quark matter \cite{Collins} (we do not consider the
case of pure quark stars, see e.g.~\cite{Itoh:1970uw}). Therefore
direct Urca processes involving quarks, i.e. the processes $d \to
u + e^- +\bar\nu_e$ and $u + e^- \to d +\nu_e$, may take place and
largely contribute to the cooling of the star. It has been shown
by Iwamoto \cite{Iwa} that quark direct Urca processes are
kinematically allowed and the corresponding emission rate for
massless quarks is of the order $\alpha_s T^6$, where $\alpha_s$
is the strong coupling constant. This result assumes that quark
matter is a normal Fermi liquid. However, since the temperature of
aged compact stars is sufficiently low, deconfined quarks in the
stellar core are likely to form Cooper pairs and quark matter
could be in one of the possible Color Superconductive (CS) phases
(see \cite{Alford:1997zt}, \cite{Alford:1998mk} and for reviews
\cite{reviews}). This is due to the fact that the critical
temperature of CS matter is of the order of dozens  MeV, well
above the estimated temperature of the stellar core $\lesssim 100$
KeV.

At asymptotically high densities the energetically favored CS phase
is the color-flavor locked (CFL) phase, in which light quarks of any
color
 form Cooper pairs with zero total
momentum \cite{Alford:1998mk}, and all fermionic excitations are
gapped. The corresponding neutrino emissivity and  specific heat $C$
are suppressed by a factor $e^{-\Delta/T}$, where $\Delta$ is the
quasiparticle gap in the CFL phase.  Therefore the cooling of quark
matter in the  CFL phase  is distinctly less rapid than in the
normal phase.
 In the CFL phase quark masses can be neglected and color and electric neutrality conditions are
 automatically implemented.
 However at densities relevant
for compact stars the quark number chemical potential $\mu$ cannot
be much larger than 500 MeV and effects due to  the strange quark
mass $m_s$ must be included. Requiring that bulk quark matter is
in weak equilibrium and electrically and color
neutral~\cite{Iida:2000ha,Amore:2001uf,Alford:2002kj,Steiner:2002gx,Huang:2002zd},
together with $m_s\neq 0$, determines a  mismatch $\delta\mu$
between the Fermi momenta of different quarks, with $\delta\mu$
depending on the in-medium value of $m_s$. For values of $m_s$
less than a critical value, CFL is the energetically favored
phase. For larger values the CFL phase cannot be realized and
quark matter should pair with a less symmetric pattern.

The ground state of quark matter in these conditions is still a
matter of debate and several possible superconductive phases have
been suggested (see \cite{Alford:2006fw} for a review). Recently two
 superconductive phases characterized by gapless fermionic excitations,
i.e. the gapless-2SC (g2SC) phase \cite{Shovkovy:2003uu}  and the
gapless-CFL (gCFL) phase
\cite{Alford:2003fq,Alford:2004hz,Alford:2004nf} have been largely
discussed. However, it has been shown
\cite{Huang:2004bg,Casalbuoni:2004tb,Fukushima:2005cm,Alford:2005qw}
 that both  are
``chromo-magnetically unstable" because the Meissner masses of some
of the gluons associated with broken gauge symmetries are imaginary.

Another possibility that has attracted theoretical attention is
the Larkin-Ovchinnikov-Fulde-Ferrell (LOFF) state
\cite{LOFF,Alford:2000ze,Bowers:2002xr,Casalbuoni:2003wh}, where
the total momentum of the pair does not vanish and counter
propagating color currents are spontaneously generated. A
simplified ansatz ``crystal" structure with
\begin{equation}
\langle\psi_{\alpha i}(x)C \gamma_5 \psi_{\beta j}(x) \rangle
\propto \sum_{I=1}^3 \Delta_I\, e^{2i{\bf q_I\cdot
r}}\epsilon_{\alpha\beta I}\epsilon _{i j I}\ , \label{cond}
\end{equation}
($i,j=1,2,3$ flavor indices, $\alpha,\beta=1,2,3$ color indices) has
been studied  in Refs. \cite{Casalbuoni:2005zp,Mannarelli:2006fy}
and  found energetically favored with respect to the gCFL and the
 unpaired phases in a certain range of values of $\delta\mu$.
In Eq. \eqref{cond} $2\,{\bf q_I}$ represents the  momentum of the
Cooper pair and  the gap parameters $\Delta_{1}$, $\Delta_{2}$,
$\Delta_{3}$ describe respectively $d-s$, $u-s$ and $u-d$ pairing.
For sufficiently large $\mu$ the energetically favored phase is
characterized by $\Delta_1=0$, $\Delta_2=\Delta_3$ and $\bf q_2 =
q_3$. Gluon Meissner masses corresponding to broken color
generators have been evaluated in \cite{Ciminale:2006sm} and this
phase results to be chromo-magnetically stable. More complex
crystal structures have been recently proposed in
\cite{Rajagopal:2006ig}. From the evaluation of the corresponding
free-energy in the Ginzburg-Landau approximation one finds  that
crystal structures with more plane waves are energetically favored
with respect to the normal phase and the gCFL phase in a wider
range of densities. We note that by (1) we assume attractiveness
in the color antisymmetric channel. This follows from the one
gluon exchange diagram of QCD. It dominates the asymptotic regime
and we assume that it favors attraction also at moderate densities
($\mu\simeq 500$ MeV). In other words, though at moderate density
nonperturbative effects can play a role we are assuming that the
superconductive ground state is qualitatively similar to that of
very high density.

The aim of the present paper is to evaluate the neutrino emission
rate and the specific heat of quark matter in the LOFF
superconductive phase. We show that, due to the existence of
gapless modes in the LOFF phase, a neutron star with a quark LOFF
core cools faster than a star made by nuclear matter only. This
follows from the fact that in the LOFF phase neutrino emissivity
and quark specific heat are parametrically similar to the case of
unpaired quark matter ($\varepsilon_{\nu} \sim T^6$ and $C\sim T $
respectively). Therefore the cooling is similar to that of a star
comprising unpaired quark matter. Incidentally we note that the
dominance of quasiparticle gapless modes has been demonstrated
also for the g2SC phase \cite{Schafer:2004jp} and for the  gCFL
phase \cite{Alford:2004zr}, though in the latter phase the
behavior of neutrino emissivity and quark specific heat is
different ($\varepsilon_{\nu} \sim T^{5.5}$ and $C\sim T^{0.5} $).
For calculation of neutrino emissivity in other models
see~\cite{Jaikumar:2005hy}.

Our results should be considered as preliminary, since, as we
 have already pointed out, the simple ansatz (\ref{cond})
 should be substituted by a more complex behavior as in
 \cite{Rajagopal:2006ig}. If the condensate is the sum of
 more plane waves, the calculation of $\varepsilon_\nu$ and $C$ is more
 involved, basically because one should obtain
 information on the quasiparticle dispersion law in the framework of
 the Ginzburg Landau
 expansion. For complex condensate patterns resulting from several plane waves this is a complicated task
 (a calculation for two flavors is in
 \cite{Casalbuoni:2003sa}) and we leave it as a future work.
Given these limitations it would be fruitless to consider
sophisticated star models. Therefore we evaluate the cooling rate by
employing toy models of stellar objects, i.e. stars made of nuclear
matter with a core in the color superconductive LOFF phase. Also, we
will not consider the contribution to the cooling of the compact
star due to other processes such as the neutrino pair bremsstrahlung
from nuclei in the crust and pionic reactions \cite{BahcallWolf}
that should also be included in a realistic description of the
compact star cooling \cite{ShapiroTeukolsky}.

Our paper is organized as follows. In  Section \ref{LOFF} we briefly
discuss the LOFF phase characterized by the condensate (\ref{cond}).
In Section \ref{Emissivity} we evaluate the neutrino emissivity  and
in Section \ref{specificheat} the specific heat in the LOFF phase.
In Section \ref{Sec:1SUmu} we evaluate the effect of $1/\mu$
corrections, taking into account the results of
\cite{Casalbuoni:2006zs}. In Section \ref{cooling} we estimate the
cooling rate of toy models of neutron stars. Given our
approximations any comparison with experimental observation, similar
e.g. to those already appeared in the literature
\cite{Blaschke:2000dy,Page:2000wt,Prakash:2000jr,Slane:2002ta,kaplan,
halpern,Page:2005fq}, is premature and we limit our analysis to a
qualitative comparison among various simple models. In Section
\ref{conclusioni} we draw our conclusions and in the Appendix we
report the dispersion laws that are used to compute the quark mixing
coefficients and gapless points in the LOFF phase.

\section{Neutral LOFF quark matter\label{LOFF}}
Non interacting quark matter consisting of massless $u$ and $d$
quarks and $s$ quarks with an in-medium mass $m_s$ can be described
by the Lagrangian density
\begin{equation}
{\cal L}_0=\bar{\psi}_{i\alpha}\,\left(i\,\de\!\!\!
/^{\,\,\alpha\beta}_{\,\,ij} -M_{ij}^{\alpha\beta}+
\mu^{\alpha\beta}_{ij} \,\gamma_0\right)\,\psi_{\beta j}
\label{lagr1}\ \,,
\end{equation}
where $i,j=1,2,3$ are flavor indices and $\alpha,\beta=1,2,3$ are
color indices; Dirac indices have been suppressed; the mass matrix
is given by $M_{ij}^{\alpha\beta} =\delta^{\alpha\beta}\, {\rm
diag}(0,0,m_s)_{ij} $ and
$\de^{\alpha\beta}_{ij}=\partial\delta^{\alpha\beta}\delta_{ij}$.
The quark chemical potential matrix is as follows:
\begin{equation}\mu^{\alpha\beta}_{ij}=(\mu\delta_{ij}-\mu_e
Q_{ij})\delta^{\alpha\beta} + \delta_{ij} \left(\mu_3
T_3^{\alpha\beta}+\frac{2}{\sqrt 3}\mu_8 T_8^{\alpha\beta}\right) \,
, \label{mu}
\end{equation} with  $Q_{ij} = {\rm
diag}(2/3,-1/3,-1/3)_{ij} $ the quark electric-charge matrix and
$T_3$ and $T_8$ the Gell-Mann matrices in color space. We are
interested in the region close to the second order phase transition
point between the normal phase and the Color Superconductive phase,
where, to the leading order approximation in $\delta\mu/\mu$,
$\mu_3=\mu_8=0$ and $\mu_e=m_s^2/4\mu$ as in the unpaired
phase~\cite{Casalbuoni:2006zs}. In all the numerical estimates we
use the value  $\mu=500$~MeV.

In order to describe quark interaction we employ a Nambu-Jona
Lasinio (NJL) model in a mean field approximation. The use of the
NJL model can be motivated by renormalization group analyses
\cite{Evans:1998nf} showing the dominance of the four-fermion
interactions and in particular the numerical importance of the
coupling mimicking one gluon exchange. We assume here that these
results hold both at high and moderate hadronic densities. The
resulting quark condensate is antisymmetric in color and flavor
indices and we assume the behavior \eqref{cond} for it. This phase
has been studied employing a Ginzburg-Landau (GL) expansion in
Ref.~\cite{Casalbuoni:2005zp}. Requiring color and electric
neutrality results in $\Delta_1=0$ and, to the leading order in
$1/\mu$ expansion, $\Delta_2=\Delta_3$. In
\cite{Casalbuoni:2006zs} it has been shown that such corrections
affect this result so that $\Delta_2\,<\,\Delta_3$. These effects
will be discussed in Section \ref{Sec:1SUmu} while in this Section
we take into account only the leading order effect in
$\delta\mu/\mu$.

The  GL approximation is reliable in a region close to the second
order phase transition point \cite{Mannarelli:2006fy} where the
favored ``crystal" structure is characterized by $\bf q_2$=$\bf
q_3$=$\bf q$. The LOFF phase is energetically favored with respect
to the gCFL phase and to the normal phase in the range of the
chemical potential mismatch $y\in (130,150)$ MeV, with\be
y=\frac{m^2_s}{\mu}\,.\label{eq4}\ee Moreover the magnitude of $\bf
q$ is given by  \be q \simeq \frac y{8\,z_q}~,~~~~~z_q=\frac
1{1.1997}\, ,\ee and $\Delta=\Delta_2=\Delta_3\le 0.3\Delta_0$ where
$\Delta_0$ is the value of the gap in the two-flavor homogeneous
case \cite{Rajagopal:2006ig}. We use the value $\Delta_0=25$ MeV.
 Recent analysis
\cite{Rajagopal:2006ig} have shown that the range of values of $y$
where the LOFF state prevails can be enlarged assuming a more
complex pattern of space dependence for $\Delta_I({\bf r})$. For the
purposes of this paper it is however sufficient to consider the
simple ansatz \eqref{cond}.

In the basis $A=(1,...,9)=
(u_r,\,d_g,\,s_b,\,d_r,\,u_g,\,s_r,\,u_b,\,s_g,\,d_b) $ the quark
chemical potentials  can be expressed by \be\mu_A=\mu+\bar\mu_A\ee
with \begin{equation}
  \bar\mu_1=-\frac{y}{6}-2\,q\,z\,,~~~~~\bar\mu_2=\frac{y}{12}\,,
 ~~~~~\bar\mu_3=-\frac{5y}{12}\,,~~~~~
  \bar\mu_4=+\frac{y}{12}-q\,z\,,~~~~~\bar\mu_5=-
  \frac{y}{6}+q\,z\,,\,\,\,\,\,
\end{equation}
\begin{equation}
 \bar\mu_6=-\frac{5y}{12}-q\,z\,,~~~~~\bar\mu_7=-\frac{y}{6}+q\,z\,,~~~~~\bar\mu_8=-\frac{5y}{12}
  \,,~~~~~\bar\mu_9=\frac{y}{12}\,.
\end{equation} Here \be z=\cos\vartheta~,\ee with $\vartheta$
the angle between the quark momentum and the pair momentum. For
later convenience we also define \be\mu_A(0)=\mu_A(q=0)\ .\ee We
note explicitly that the previous chemical potentials can be
obtained by a redefinition of the quark fields in the following way:
\be \tilde\psi_{\alpha i}(x)=e^{i{\bf q}_{\alpha i}\cdot{\bf
x}}\psi_{\alpha i}(x)\,,\label{redef0}\ee where ${\bf q}_{\alpha
i}={\bf q}$ for $(\alpha, i)=[(r,d),(g,u),(r,s),(b,u)]$, ${\bf
q}_{\alpha i}=2{\bf q}$ for $(\alpha, i)=(r,u)$ and  ${\bf
q}_{\alpha i}=0$ in the other cases. Such a  redefinition of the
quark fields allows to eliminate the space dependence of the
condensate. However quark momenta are shifted as follows: \bea
&&~~~~~~~~~~~~~~~~~~ {\bf p}_{u_r}\rightarrow {\bf p}_{u_r}-2\,{\bf
q}~, ~~~~{\bf p}_{d_g}\rightarrow {\bf p}_{d_g}~,~~~~{\bf
p}_{s_b}\rightarrow {\bf p}_{s_b}~,\cr &&{\bf p}_{d_r}\rightarrow
{\bf p}_{d_r}-{\bf q}~, ~~~~{\bf p}_{u_g}\rightarrow {\bf
p}_{u_g}-{\bf q}~,~~~~{\bf p}_{s_r}\rightarrow {\bf p}_{s_r}-{\bf
q}~,~~~~~{\bf p}_{u_b}\rightarrow {\bf p}_{u_b}-{\bf q}~,\cr&&
~~~~~~~~~~~~~~~~~~~~~~~~~~~~~~~~~ ~~~{\bf p}_{s_g}\rightarrow {\bf
p}_{s_g} ~,~~~~{\bf p}_{d_b}\rightarrow {\bf p}_{d_b}\
.\label{redef}\eea
\section{Neutrino emissivity \label{Emissivity}}The
transition rate for  the $\beta$ decay of a down quark  $d_\al$, of
color $\al=r,g,b$, into an up quark $u_\al$ \be d_\al(p_1)~\to~
\bar\nu_e(p_2)~+~u_\al(p_3)~+~ e^-(p_4)\label{process}\ee
 is \be
W_{fi}=V(2\pi)^4\delta^4(p_1-p_2-p_3-p_4)|\mathcal{M}|^2\prod_{i=1}^4
\frac{1}{2E_iV}~,\ee where $V$ is the available volume and
$\mathcal{M}$ is the invariant amplitude. Neglecting quark masses
the squared invariant amplitude averaged over the initial spins and
summed over spins in the final state
is\begin{equation}\label{eq:feyn_ampl}
  |\mathcal{M}|^2=64 G^2_F \cos^2\theta_c (p_1 \cdot p_2)(p_3 \cdot
  p_4)\,,
\end{equation}
where $G_F$ is the Fermi constant and $\theta_c$ the Cabibbo angle;
we will neglect the strange-quark $\beta$ decay whose contribution
is smaller by a factor of $\tan^2\theta_c$ in comparison with
\eqref{eq:feyn_ampl}.  For relatively aging stars, there is no
neutrino trapping~\cite{Ruster:2005ib} which means that  neutrino
momentum and energy are small. The magnitude of the other momenta is
of the order of the corresponding Fermi momenta $p_{F}^1$, $p_{F}^3$
and $p_{F}^4$. As discussed in Section \ref{LOFF}, for $d$ and $u$
quarks they are of the order of 500 MeV; the electron momentum is of
the order $\mu_e$, which is smaller, but, due to the assumptions
discussed in the previous section, still sizeable. On the other hand
the neutrino momentum $p_2$ is of the order $k_BT$. It follows that
the momentum conservation can be implemented neglecting $\bf {p_2}$
and one can depict the 3-momentum conservation for the decay
\eqref{process} as a triangle \cite{Iwa} having for sides ${\bf
p_1}$, ${\bf p_3}$ and ${\bf p_4}$.  It follows that we can
approximate \be (p_1 \cdot p_2)(p_3 \cdot
  p_4)\simeq E_1E_2E_3E_4(1-\cos\theta_{12})
  (1-\cos\theta_{34})\ee where $E_j$ are the energies and  $\theta_{12}$ (resp. $\theta_{34}$)
  is the  angle between momenta of the  down quark  and the neutrino
  (resp. between the up quark and the electron).
  Neutrino emissivity $\varepsilon_{\nu}$ is defined as the energy loss by
  neutrino emission per second per unit volume. To compute it,
  we have to multiply
  \eqref{eq:feyn_ampl} by the neutrino energy and thermal
  factors, integrate over the available phase space and sum over the three colors \cite{Iwa}. Moreover, since the states having definite energy are the quasiquarks
  involved in the Cooper pairs, one has to multiply the result by
  two mixing (Bogoliubov) coefficients depending on the quasiparticle
  dispersion laws \cite{Alford:2004zr}. Therefore we get\begin{eqnarray}
\varepsilon_{\nu}&=&\sum_{\al=r,g,b}\varepsilon_{\nu}^\alpha=\sum_{\al=r,g,b}
\frac{2}{V}\left[\prod\limits_{i=1}^4
\int\frac{d^3p_i}{(2\pi)^3}\right] E_2 W_{fi}\nn\\
&&\cdot n({\bf p_{1}})\cdot[1-n({\bf p_{3}})]\cdot[1-n({\bf
p_4})]\cdot B_{d_\alpha }^2({\bf p_{1}})B_{ u_\alpha}^2({\bf p_{3}})
\label{eq:emissDef}
\end{eqnarray}where $n({\bf p_{j}})$ are thermal equilibrium Fermi
distributions: \be n({\bf p_j})=\frac{1}{1+\exp{x_j}}~,\ee\be
x_j=\frac{E_j({\bf p_j})-\mu_j}T
\ .\label{x}\ee
They are appropriate here because strong
and electromagnetic processes establishing thermal equilibrium are
much faster than weak interactions. The overall factor of $2$ keeps
into account the electron capture process.

Thus far the results are similar to those of \cite{Alford:2004zr}.
However, in the present case, the  phase space integrations  and
the momentum conservation law must be treated with great care. As
a matter of fact, in the LOFF color superconductive state $d_\al$
and $u_\al$ are in general paired with quarks of another color,
and there is breaking of translational invariance because the
total momentum of the Cooper pair is $2\bf q$. Let us  choose the
$z-$axis along the direction of $\bf q$ and let $\vartheta_j$ and
$\phi_j$ be the polar angles of $ \bf p_j$. Since the temperature
is much smaller than the gap parameter, the dominant modes in the
$d-$ and $u-$momentum integrations are the gapless ones, that we
denote as $P_j$ (with $j=1$ for down quark and $j=3$ for up
quark). They are defined as follows: $P_1$ (resp. $P_3$) is the
quark down (resp. quark up) momentum where the corresponding
quasi-particle energy vanishes (for more details see the
Appendix). On the other hand the relevant momentum for the
electron is its Fermi momentum. Therefore we have
  \be\int d^3p_1\int d^3p_3\int d^3p_4~\approx ~\int\mu_e^2 dp_4d\Omega_4~ P^2_{1}
  ~dp_1~d\Omega_1P^2_{3}~dp_3~d\Omega_3~,\ee with $d\Omega_j=\sin\vartheta_j\,
  d\vartheta_j~d\phi_j$. While these approximations are similar to
  the assumptions used in \cite{Alford:2004zr} in dealing with the homogeneous gCFL phase,
  in the LOFF phase  the gapless momenta $P_1$ and $P_3$ depend on
  the angle $\vartheta_j$  that quark momenta form with the pair momentum $2\,\bf q$.
   In order to simplify the expression of the integral in Eq.~\eqref{eq:emissDef} the  variables $p_j$ can be traded for
   $x_j$, as in Eq.~\eqref{x}. We put $x_2=E_2/T$ for the neutrino,
     $x_4=(p-\mu_e)/T$ for the electron and $ E_j=\mu_j+\epsilon_j(p) $ for the quarks.
     Expanding  around the gapless modes one has
    $E_j(p)\simeq \mu_j+v_j\,(p-P_j)$ (for $ j=1$ and $3$), with
the quasiparticle velocity given by \be\label{velocity}
  v_j=\left.\frac{\partial E_j}{\partial
  p}\right|_{p=P_j}~.
\ee As discussed in detail in the Appendix the dispersion law of
each quasiparticles has from one to three gapless modes. Therefore
one has to expand the corresponding dispersion laws around each
gapless momentum. Another point to be stressed concerns the
conservation of three-momentum. As we have noted, to get rid of the
space dependence in the condensate, one redefines the quark fields,
see Eq.~\eqref{redef0}, which amounts to a redefinition of momenta
as in Eq.~\eqref{redef}.  Using the new quark fields in the weak
decay matrix element \eqref{eq:feyn_ampl} adds a momentum $-\bf q$
to the momentum conservation law that now reads: ${\bf p_1}-{\bf
p_3}-{\bf p_4}-{\bf q}=0$ (where the neutrino momentum has been
neglected).

Employing the above approximations the neutrino emissivity for each
pair of gapless momenta $P_1,P_3$, can be written as
\bea\varepsilon_\nu^\alpha&\simeq&\frac{G_F^2\,\cos^2\theta_c\,\mu_e^2\,T^6}{32\pi^8}
\,I\,\int\frac{ d{\bf r}}{(2\pi)^3} \prod_{j=1}^4\int
d\Omega_j~\frac{P_1^2P_3^2\,B_{d_\alpha }^2({P_{1}})B_{ u_\alpha}^2
({P_{3}})}{|v_1|\,|v_3|}(1-\cos\theta_{12})
  (1-\cos\theta_{34})~e^{-i\bf{r\cdot}({\bf p_1}-{\bf p_3}-{\bf p_4}-{\bf q})}
  \cr
  &=&\frac{G_F^2\,\cos^2\theta_c\,\mu_e^2\,T^6}{8\pi^7}
\,I\,\int\frac{ d{\bf r}}{(2\pi)^3} \prod_{j=1,3,4}\int
d\Omega_j~\frac{P_1^2P_3^2\,B_{d_\alpha }^2({P_{1}})B_{ u_\alpha}^2
({P_{3}})}{|v_1|\,|v_3|}\cr&&\times
  \left(1-\frac{P^2_1+q^2-2qP_1\cos\vartheta_1-P^2_3-\mu_e^2}{2\,\mu_e\,P_3}
  \right)~e^{-i\bf{r\cdot}({\bf p_1}-{\bf p_3}-{\bf p_4}-{\bf q})}\label{emissivity} \, ,\eea
where  $P_1$ and $P_3$ depend, respectively, on the angles
$\vartheta_1$ and $\vartheta_3$ and $B_{i_\alpha }$ is a mixing
coefficient arising from the Bogoliubov-Valatin transformation
representing the probability amplitude that the gapless
quasiparticle has flavor $i$ and color $\alpha$. Moreover
\cite{Iwa,Morel}\be
  I=
  \int_{-\infty}^{+\infty} dx_1\int_{0}^{+\infty}
   dx_2~x_2^3\int_{-\infty}^{+\infty} dx_3
\int_{-\infty}^{+\infty} dx_4 n(x_1)n(-x_3)n(-x_4)
\delta(x_1-x_2-x_3-x_4)~=~\frac{457\pi^6}{5040}\,. \ee Some of the
angular integrations appearing in \eqref{emissivity}
 can be performed analytically, with the result
\begin{eqnarray}
 \varepsilon_\nu^\al &\simeq&
 \frac{G_F^2\,\cos^2\theta_c\,\mu_e^2\,T^6}{4\pi^6}\,I\,
  \sum\limits_{k=1}^5
  \int\nolimits_0^{+\infty}dr\,\, r^2\int\nolimits_{-1}^{1} d(\cos\vth)e^{i\,
  qr\cos\vartheta}\cr
 &&\times\int_{\omega_0}^{\omega_1} d(\cos\vth_1) e^{-iP_1 r
\cos\vth \cos\vth_1} J_0(P_1 r \sin\vartheta\sin\vartheta_1)
f_k(P_1)\frac{P_1^2\, B_{d_\al}^2(P_1)}{|v_1|}\cr
&&\times\int_{\omega_2}^{\omega_3}d(\cos\vth_3) e^{iP_3 r \cos\vth
\cos\vth_{3}} J_0(P_3 r \sin\vartheta\sin\vartheta_3)
g_k(P_3)\frac{P_3^2\,B_{u_\al}^2(P_3)}{|v_3|}\cr
 &&\times\int_{-1}^{1} d(\cos\vth_4) e^{i\mu_e r \cos\vth
\cos\vth_4} J_0(\mu_e r \sin\vartheta\sin\vartheta_4)\ .
\label{eq:EMISSIVITY_f}\end{eqnarray}Here for convenience we have
decomposed the integral as a sum over various terms, with
$f_1=f_3=f_4=1$~,~ $f_2=\,P_1^2+q^2$,
~$f_5=\,-\,2P_1q\cos\vartheta_1$, and $g_1=1$,
   $\displaystyle g_2=g_5=~\,-\,\frac 1{2\mu_e P_3}~, ~g_3=\frac{P_3}{2\mu_e}~,~
    g_4=\frac{\mu_e}{2P_3}~$.

In Eq.~\eqref{eq:EMISSIVITY_f}  $J_0$ is the Bessel function of
zeroth order; $\omega_0$, $\omega_1$, $\omega_2$, $\omega_3$ are
appropriate limits taking into account kinematic constraints;
$\vartheta$ is the angle between the directions of $\bm r$ and $\bm
q$. Even if each quasiparticle dispersion law is characterized by
various gapless momenta, the number of the  gapless momenta relevant
for the evaluation of the neutrino emissivity  can be reduced
observing that the momenta $\bf p_1$, $\bf p_3$ and $\bf p_4$, with
$\bf |p_4| \simeq \mu_e$, must satisfy the condition  ${\bf
p_1}-{\bf p_3}-{\bf p_4}-{\bf q}=0$.

We now consider the contributions of the three colors separately. We
discuss in detail only the decay of the blue quark that also gives
the largest contribution. The decay of the other colors is treated
analogously. The blue down quark is unpaired because in the
Ginzburg-Landau approximation the gap parameter $\Delta_1$ vanishes
\cite{Casalbuoni:2005zp}. Therefore, as discussed in the Appendix,
the dispersion law is $E_1(p)=p$, with
 gapless momentum $P_1=\mu_9$, mixing coefficient $B_{d_b}(P_1)=1$
 and $v_1=1$.

Let us now consider the up-blue quark. It is paired with the
strange-red quark with gap parameter $\Delta_2=\Delta$. In the
Appendix (subsection \ref{sett67}) we give the dispersion law and
the gapless momentum: \be P_{3}= \frac{\mu_6+\mu_7}2+
 \sqrt{\frac{y^2}{64}
  \left(1+\frac{\cos\vartheta_3}{z_q}\right)^2-\D^2}\,.\label{gap3}
\ee We do not consider the momentum corresponding to the other
gapless mode of Eq. \eqref{gapless_momenta_67}, as it does not
satisfy the condition of momentum conservation. The emissivity for
the decay of the blue quark is therefore:
\begin{eqnarray}
 \nn \varepsilon_\nu^{\rm blue} &\simeq&
 \frac{G_F^2\,\cos^2\theta_c\,\mu_e^2\mu_9^2\,T^6}{4\pi^6}\,I\,
    \int_0^{+\infty}dr\,\, r^2\int\nolimits_{-1}^{1} d(\cos\vth)e^{iqr\cos\vartheta}\\
\nn &&\times\int\nolimits_{-1}^{+1} d(\cos\vth_1) e^{-iP_1 r
\cos\vth \cos\vth_1} J_0(P_1 r \sin\vartheta\sin\vartheta_1)
\\
\nn &&\times\int_{\omega_2}^{+1}d(\cos\vth_3) e^{iP_3 r \cos\vth
\cos\vth_{3}} J_0(P_3 r \sin\vartheta\sin\vartheta_3)
g(\cos\vartheta_3\,,\cos\vartheta_1)\\
 &&\times\int_{-1}^{1} d(\cos\vth_4) e^{i\mu_e r \cos\vth
\cos\vth_4} J_0(\mu_e r \sin\vartheta\sin\vartheta_4)
\label{eq:EMISSIVITYblue}\, ,\end{eqnarray} where  \be
\omega_2={\text{
Min}}\left\{z_q\left[\left(\frac{8\Delta}{y}\right)-1\right]~,~+1\right\}~,\label{omega2}\ee
\bea
g(\cos\vartheta_3\,,\cos\vartheta_1)&=&\left(1-\frac{\mu_d^2+q^2-2q\mu_d\cos\vartheta_1-P_3^2
-\mu_e^2}{2P_3\mu_e}\right)\frac{\left[P_3\,B_{u_b}(P_3)\right]^2}{|v_3|}\cr&\times&\Theta\left(1-\left|\frac{\mu_d^2+q^2-2q\mu_d\cos\vartheta_1-P_3^2
-\mu_e^2}{2P_3\mu_e}\right|\right)\eea and \be
 |v_{3}|=\frac{\displaystyle \left|P_3-
 \frac{\mu_6+\mu_7}2\right|}{\displaystyle\sqrt{\left(P_3-
 \frac{\mu_6+\mu_7}2\right)^2+\Delta^2}}~~.\label{vutre}
\ee  We note that the presence of $\omega_2$ implements the
existence of the gapless momentum $P_3$ while the Heaviside function
implements the condition $|\cos\vartheta_3|\le 1$.

 The case of quarks with colors red and green can be treated in a similar way, though the
numerical computation is more involved because for these colors
neither the down nor the up quarks are unpaired. Because of this one
expects that the emissivity of these quarks is smaller than for blue
quarks, an expectation confirmed by the numerical analysis. We will
report our numerical results in Section \ref{cooling}.

\section{Specific heats \label{specificheat}}

At low temperatures the largest contribution to  specific heat $C$
is determined  by the sum of the specific heats of the fermionic
quasi-particles. In the three flavor LOFF phase  the quasi-particle
specific heats are given by \begin{equation} c_j =
2\!\int\!\frac{d^3\!p}{(2\pi)^3}~\epsilon_j \frac{\partial
n(\epsilon_j)}{\partial T}~,\label{eq:CvDef}
\end{equation}
where $j=u_r\,,d_g\,,s_b\,, d_r\,, u_g\,, s_r\,, u_b\,, s_g\,, d_b$
and $\epsilon_j=\left|E_j({\bf p})-\mu_j\right|$ are the
quasi-particle dispersion laws. Since we work in the regime
$T\ll\Delta\ll\mu$, the contributions of gapped modes are
exponentially suppressed and each gapless mode  contributes by a
factor $\propto T$. This results follows from the evaluation of the
integral in Eq.(\ref{eq:CvDef}) employing  the saddle point method
and assuming that the quasi-particle dispersion laws are linear in
the gapless momenta. For the present choice of parameters, the
quasiparticle dispersion law is quadratic in the gapless momenta in
a negligible range of values of the angle between the direction of
quasiparticle momentum and the direction of $\bf q$. Such a
situation is quite different from the homogeneous gCFL case where a
quasiparticle dispersion law is quadratic \cite{Alford:2004zr} and
gives the dominant contribution to   the specific heat.

Within the above-mentioned approximations  the fermionic  specific
heath is given by  \be C  \simeq\frac{T}{3}
\int\frac{d\cos\vartheta}2\,\Big[\frac{P_I^2}{|v_I|}+\frac{P_{II}^2}{|v_{II}|}
+\frac{P_{III}^2}{|v_{III}|}
+\frac{(P_+^{u_gd_r})^2}{|v_+^{u_gd_r}|}+\frac{(P_-^{u_gd_r})^2}{|v_-^{u_gd_r}|}+
\frac{(P_+^{u_bs_r})^2}{|v_+^{u_bs_r}|}+\frac{(P_-^{u_bs_r})^2}{|v_-^{u_bs_r}|}
+(P_-^{d_b})^2+(P_-^{s_g})^2+\mu_e^2\Big] \label{c}  \ee where the
various gapless momenta and Fermi velocities on the r.h.s have been
defined in the Appendix and in Eq. \eqref{velocity}. In particular
$P_{I}$, $P_{II}$ and $P_{III}$ are the gapless momenta of the
sector $(u_r\,,d_g\,,s_b)$ and  $v_{I}$, $v_{II}$ and $v_{III}$ the
corresponding velocities. We have also added the electron
contribution, though $\mu_e$ is numerically much smaller than all
the quark momenta $P_g$ (generally of the order of $\mu$). We also
note that the integration range is in general smaller than $(-1,+1)$
because the quasi-particle dispersion laws are gapless only in a
restricted range of values of the variable $\cos\vartheta$.

\section{Role of $1/\mu$ corrections}\label{Sec:1SUmu}
In previous sections we have evaluated emissivity and specific heat
in the three flavor LOFF phase of QCD. All the calculations are
based on the High Density Effective Theory, see
\cite{Hong:1998tn,Beane:2000ms,Casalbuoni:2003cs}, where one takes
the limit $\mu\rightarrow\infty$. In such approximation one neglects
the contribution of the antiparticles; moreover the $m_s\neq 0$
effects are treated at the leading order by a shift in the strange
Fermi momentum $p^F_s\approx\mu_s-m_s^2/2\mu$. One can show
\cite{Casalbuoni:2006zs} that, as already stressed in Section
\ref{LOFF}, in this approximation the electron chemical potential
$\mu_e$ is given by $\mu_e= m_s^2/4\mu$, which implies a symmetric
splitting of the $s$ and $d$ Fermi surfaces around the $u$ Fermi
surface. Therefore $\Delta_1=0$ and $\Delta_2=\Delta_3$. If $\mu$ is
not large enough this approximation is not justified and higher
order corrections must be included

The next-to-leading corrections were computed in
\cite{Casalbuoni:2006zs}. One expands the strange quark momentum up
to the next-leading order:
\begin{equation}
p^F_s \approx \mu_s - \frac{m_s^2}{2\mu_s} -
\frac{1}{2\mu}\left(\frac{m_s^2}{2\mu}\right)^2  \label{eq:Pfs}\,.
\end{equation}
Then one substitutes this expression in the Lagrangian in
Eq.(\ref{lagr1}), getting a correction. It can be proven
\cite{Casalbuoni:2006zs} that other corrections, e.g. the
antiparticle contribution, is next-to-next-to-leading in the weak
coupling approximation and therefore can be safely neglected.

Keeping only terms of ${\cal O}(\Delta/q)^4$ in the free energy, the
$1/\mu$ shift in Eq.~\eqref{eq:Pfs} results in the chemical
potentials
\begin{equation}
\mu_e = \frac{m_s^2}{4\mu} -
\frac{m_s^4}{48\mu^3}~,~~~~~\mu_3,~\mu_8 = 0 ~.
\label{eq:NewChemPot}
\end{equation}
The introduction of the $1/\mu$ terms in the free energy results in
an asymmetric splitting of the Fermi surfaces of the $u$, $d$ and
$s$ quarks. As a matter of fact one gets in the normal phase
\begin{equation}
\frac{\mu_u - p^F_s}{2} = \frac{m_s^2}{8\mu} +
\frac{5}{96}\frac{m_s^4}{\mu^3}~,~~~~~~~~~~ \frac{\mu_d - \mu_u}{2}
= \frac{m_s^2}{8\mu} - \frac{1}{96}\frac{m_s^4}{\mu^3}~.
\end{equation}
As a consequence one expects in the LOFF phase still $\Delta_1=0$,
but $\Delta_2< \Delta_3$. For example for $\mu = 500$ MeV one finds
at
 $m_s^2/\mu = 140$ MeV  that $\Delta_2
= 0$ and $\Delta_3 \simeq 0.35\Delta_0$. Thus for this value of
$m_s$ the vacuum consists of LOFF pairs of $u_r - d_g$ and $u_g -
d_r$ quarks and Fermi seas of unpaired $s$, $u_b$ and $d_b$ quarks.
This new phase was called LOFF2s in~\cite{Casalbuoni:2006zs} since
it is a two flavor LOFF phase whose excitation spectrum consists of
quasi-particles with dispersion laws
\begin{equation}
\epsilon = \frac{\mu_4 - \mu_5}{2} \pm \sqrt{\left(p-\frac{\mu_4 +
\mu_5}{2}\right)^2+\Delta_3^2}
\end{equation}
and a sea of unpaired blue  and strange quarks.

 Motivated by these
results we study neutrino emissivity and specific heat of the LOFF2s
phase. Since blue up and down quarks are unpaired, their
contribution to the neutrino emissivity is $1/3$ of the neutral
unpaired quark matter. Using for it the estimate of Ref.~\cite{Iwa}
one obtaines
\begin{equation}
\varepsilon_{\nu}^{\rm blue} \simeq \frac{1}{3}\times\frac{457}{630}
\frac{G_F^2\cos^2\theta_c}{\hbar^{10}c^6}\,\alpha_s\,\mu_u\,\mu_d\,\mu_e\,(k_B
T)^6~,\label{unosutre}
\end{equation}
where we have restored the correct factor of $c$, $\hbar$ and $k_B$.
We note that in this equation $\alpha_s=g^2/4\pi=4\alpha_c$ where
$g$ gives the coupling among quarks and gluons and $\alpha_c$ is the
coupling constant appearing in Ref.~\cite{Iwa}. Notice that the use
of this estimate by Iwamoto \cite{Iwa} is only indicative, because
at the densities we are considering, perturbative QCD is unreliable.
Therefore \eqref{unosutre} can provide at best an order of magnitude
estimate. Neglecting the contribution of strange quarks, the other
contribution to the emissivity is determined by red and green light
quarks and can be computed following the same lines of Section
\ref{Emissivity}. Since the LOFF pairing results in a restriction of
the available phase space, the red and green channels contribute to
the total decay rate less than the blue one, similarly to the
discussion in Section \ref{Emissivity}.

Next we turn to the specific heat, which is given by
\begin{equation}
C\simeq\frac T
3\int\frac{d\cos\vartheta}2\,\left[2\times\left(\frac{(P_+)^2}{v_+}+\frac{(P_-)^2}{v_-}\right)
+(\mu^{s_b})^2 + (\mu^{s_r})^2+(\mu^{u_b})^2
+(\mu^{d_b})^2+(\mu^{s_g})^2+\mu_e^2\right] \label{eq:c2s}
\end{equation}
where the gapless momenta $P_{\pm}$ are defined as
\begin{equation}
P_\pm = \frac{\mu_4+\mu_5}{2} \pm
\sqrt{\frac{\mu_4-\mu_5}{2}-\Delta_3^2}   \label{eq:Ppm}
\end{equation}
and the Fermi velocities are given by
\begin{equation}
v_+ = v_- =
\frac{\displaystyle\sqrt{\left(\frac{\mu_4-\mu_5}{2}\right)^2-\Delta_3^2}}
{\displaystyle\left|\frac{\mu_4-\mu_5}{2}\right|}
~.\label{eq:FermiLOFF2s}
\end{equation}
The factor $2$ in the first two addenda of the r.h.s. of
Eq.~\eqref{eq:c2s} takes into account the fact that the dispersion
laws for the $u_r-d_g$ quasiparticles are the same of the $u_g-d_r$
ones.

\section{Cooling by neutrino emission \label{cooling}}

Let us assume the presence of quark matter in the LOFF state in the
core of a compact star. This will affect the cooling process and we
now discuss these effects by comparing various models of stars along
lines similar to Ref.~\cite{Alford:2004zr}.  Given the
approximations used in the study of the LOFF phase, in particular
the use of the simple ansatz \eqref{cond} instead of more complex
space behavior of the condensate, it would be fruitless to employ
sophisticated models. Therefore we will use a simplified approach
based on the study of four different star toy models. The first
model (denoted as I) is a star consisting of noninteracting
``nuclear" matter (neutrons, protons and electrons) with mass
$M=1.4M_\odot$, radius $R=12$ km and uniform density $n=1.5\,n_0$,
where $n_0 = 0.16$ fm$^{-3}$ is the nuclear equilibrium density. The
nuclear matter is assumed to be electrically neutral and in beta
equilibrium. The second model (II) is a star containing a core of
radius $R_1=5$ km of neutral unpaired quark matter in weak
equilibrium at $\mu=500$ MeV, with a mantle of noninteracting
nuclear matter with uniform density $n$. Assuming a star mass
$M=1.4\,M_\odot$ from the solution of the Tolman-Oppenheimer-Volkov
equations one gets a
 star radius $R_2=10 $ km. Finally we discuss two simplified models of
compact stars containing a core of electric and color neutral three
flavors quark matter in the LOFF phase, with  $\mu=500$ MeV and
$m_s^2/\mu=140$ MeV and $\Delta_0=25$ MeV. Both models have a mantle
of noninteracting nuclear matter and differ by the values of the
LOFF gaps. In model III we neglect $\mathcal{O}(1/\mu)$ corrections
and we take $\Delta_1=0$ and $\Delta_2=\Delta_3= 0.25\,\Delta_0$,
the value of the gap parameters in the LOFF phase at $y=140$
MeV~\cite{Casalbuoni:2005zp}. This is the approximation discussed in
Section \ref{LOFF}. In model IV we include the $1/\mu$ correction
discussed in Section~\ref{Sec:1SUmu} and we take the values of the
gaps as in the LOFF2s model discussed in
Ref.~\cite{Casalbuoni:2006zs}, i.e. $\Delta_1=\Delta_2=0$,
$\Delta_3=0.28\, \Delta_0$ for $m_s^2/\mu=140$ MeV. Since the values
of the gaps in both LOFF models are small the radii of the star and
of the quark core do not differ appreciably from those of a star
with a core of unpaired quark matter, i.e. $R_1=5$ and $R_2=10$ km
(also in these cases $M=1.4 M_\odot$).

The main mechanisms of cooling are by neutrino emission and by
photon emission, the latter dominating at later ages. Therefore the
star cooling is governed by the following differential equation: \be
\frac{dT}{dt} = - \frac{ L_\nu+L_\gamma}{V_{nm}c_V^{nm} +
V_{qm}c_V^{qm}} = - \frac{V_{nm}\varepsilon_\nu^{nm} +
V_{qm}\varepsilon_\nu^{qm} + L_\gamma} {V_{nm}c_V^{nm} +
V_{qm}c_V^{qm}}~. \label{ANGLANI1} \ee Here $T$ is the inner
temperature at time $t$, while  $L_\nu$ and $L_\gamma$ are neutrino
and photon  luminosities, i.e. heat losses per unit time. Neutrino
luminosity is obtained multiplying the
 emissivity by the corresponding volume. Therefore one must distinguish
 between the emissivity
 $\varepsilon_\nu^{nm}$ of nuclear matter, which is present in the
 volume $V_{nm}$, from the emissivity due to quarks, if they are
 present in some volume $V_{qm}$;
 $c_V^{nm}$ and $c_V^{qm}$  denote specific heats of the two forms of hadronic matter.

For  the emissivity of  nuclear matter  $\varepsilon_\nu^{nm}$ we
use the standard value \cite{ShapiroTeukolsky}:
 \be
\varepsilon_\nu^{nm}=\left(1.2\times 10^4\, {\rm erg~cm}^{-3}{\rm
s}^{-1} \right) \left(\frac{n}{n_0}\right)^{2/3}
\left(\frac{T}{10^7~{\rm K}}\right)^8 \ee arising from the analysis
of the modified Urca processes  $n+X\to p+X+e+\bar\nu$, with $X=p$
or $n$. As for the quark contribution $\varepsilon_\nu^{qm}$, we
consider both unpaired quarks and quarks in the LOFF state,
depending on the model. The former contribution is denoted
$\varepsilon_\nu^{qm,unpaired}$ and is given by Eq.\eqref{unosutre}
multiplied by a factor of 3. The latter have been estimated in
Sections \ref{Emissivity} and \ref{Sec:1SUmu} for the two models of
LOFF quark matter. For all these stars we assume a common interior
temperature $T$, since the matter comprising compact stars is made
of good conductors ($p$, $n$ or quarks).

Let us now turn to cooling by  photon emission, the dominant process
for sufficiently old stars ($t>10^6$ years). In this case the
luminosity can be estimated in the black-body approximation as
follows:
\begin{equation} L_\gamma=4\pi R^2 \sigma T^4_s\ ,
\end{equation}
where $R$ is the radius of the star, $\sigma$ is the
Stefan-Boltzmann constant and \be T_s = (0.87\times 10^6~{\rm K})^4
\left(\frac{g_s}{10^{14} {\rm cm}/{\rm s}^2}\right)^{1/4}
\left(\frac{T}{10^8 {\rm K}}\right)^{0.55}\ ,\label{surf}\ee is the
surface temperature \cite{Gundmundsson,Page:2004fy}, with  $g_s= G_N
M/R^2$ the surface gravity.

Let us now discuss specific heats.  They are given for the LOFF
model III and for the LOFF2s model IV  by Eqs.~\eqref{c}
and~\eqref{eq:c2s} respectively. For nuclear and unpaired quark
matter the total specific heat is given by the sum of the specific
heats of the different fermionic species. They can  be approximated
by the fermionic ideal gas result \be c_V=\frac{k_B^2 T}{3\hbar^3 c}
p_F\sqrt{m^2 c^2 +(p_F)^2} \ , \ee where $m$ is the fermionic mass
and $p_F$ the Fermi momentum. For non-interacting nuclear matter,
the three species are neutrons, protons and electrons with Fermi
momenta evaluated as in neutral matter in weak equilibrium
\cite{ShapiroTeukolsky}: \bea
p_F^n&=& \left(340~{\rm MeV}\right)\left(\frac{n}{n_0}\right)^{1/3}\,,\nonumber\\
p_F^p&=&p_f^e = \left(60~{\rm
MeV}\right)\left(\frac{n}{n_0}\right)^{2/3}\ . \eea For neutral
unpaired quark matter in weak equilibrium, there are nine quark
species, with Fermi momenta  independent of color and  given by $
p_F^d=  \mu+\displaystyle\frac{m_s^2}{12\mu}$, $p_F^u =
 \mu-\displaystyle\frac{m_s^2}{6\mu}$, $p_F^s =  \mu-\displaystyle\frac{5 m_s^2}{12\mu}$.

\begin{figure}[t]
\begin{center}
\includegraphics[width=8cm]{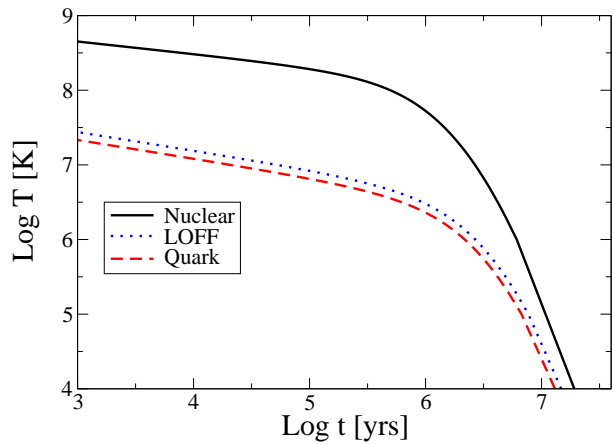}
\end{center}
\caption{ (Color online) Inner temperature, in Kelvin, as a function
of time, in years, for three toy models of pulsars. Solid black
curve  refers to model I; dashed line (red online) refers to model
II; dotted curve (blue online) refers to model III. Model I is a  a
neutron star formed by nuclear matter with uniform density $n=0.24$
fm$^{-3}$ and radius $R=12$ Km; model II corresponds to a star with
$R_2=10$ km, having a mantle of nuclear matter and a core of radius
$R_1=5$ Km of unpaired quark matter, interacting {\it via} gluon
exchange; model III is like model II, but in the core there is quark
matter in the LOFF state, see text for more details. All stars have
$M=1.4\,M_\odot$. Parameters for the core are $\mu=500$ MeV and
$m^2_s/\mu$=140 MeV.} \label{coolcore}
\end{figure}

\begin{figure}[t]
\begin{center}
\includegraphics[width=8cm]{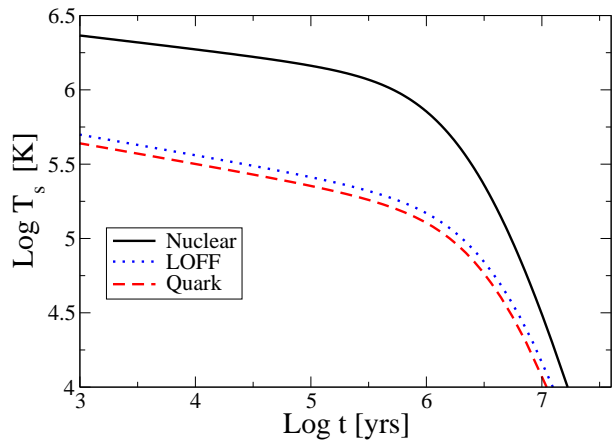}
\end{center}
\caption{ (Color online) Surface temperature $T_s$, in Kelvin, as a
function of time, in years, for the three toy models of pulsars
described in Fig. \ref{coolcore}.} \label{coolsurf}
\end{figure}

\begin{figure}[t]\begin{center}
\includegraphics[width=8cm]{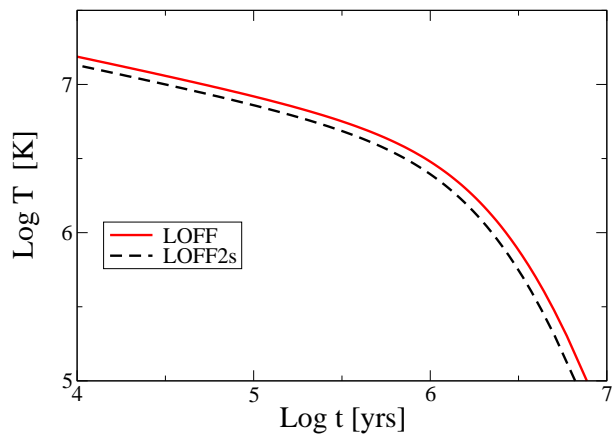}
\end{center}
\caption{ (Color online)  Comparison of the cooling curves for the
two toy models of stars with a core of quarks in a LOFF state. The
curves express the inner temperature, in Kelvin, as a function of
time, in years, for model III (solid, red online, curve) and model
IV (dashed black curve). Model III is a star with a radius $R_2=10$
km, having a mantle of nuclear matter and a core of radius $R_1=5$
Km of quark matter in the LOFF state in the leading approximation
($\Delta_1=0,\Delta_2=\Delta_3\simeq 6$ MeV); model IV is analogous
to model III, but the core is in the LOFF2s phase
($\Delta_1=\Delta_2=0\,,\Delta_3\simeq 7$ MeV). All stars have
$M=1.4\,M_\odot$. Parameters for the core are $\mu=500$ MeV and
$m^2_s/\mu$=140 MeV.} \label{coolcomparison}
\end{figure}

Eq. \eqref{ANGLANI1} is solved imposing a given temperature $T_0$ at
a fixed early time $t_0$ (we use $T_0\to\infty$ for  $t_0\to 0$). In
Figs. \ref{coolcore}, \ref{coolsurf} and \ref{coolcomparison} the
cooling curves for various models of stars are shown. In Fig.
\ref{coolcore} the inner temperature, in Kelvin, as a function of
time, in years, for three models is reported. Solid line (black
online) is for model I (electrically neutral nuclear matter made of
non interacting neutrons, protons and electrons in beta
equilibrium); dashed curve (red online) refers to model II (nuclear
matter mantle and a core of unpaired quark matter, interacting {\it
via} gluon exchange); the dotted line (blue online) is for model III
(nuclear matter mantle and a core of quark matter in the LOFF state
in the leading $1/\mu$ approximation, as discussed in Section
\ref{LOFF}). The curve reported in Figs. \ref{coolcore} and
\ref{coolsurf} for  model III corresponds to $m_s=\sqrt{140\,\mu}$
MeV, however with increasing values of $m_s$ the neutrino emissivity
decreases. This is due to the fact that $\Delta$ decreases as one
approaches the second order phase transition to the normal state. On
the other hand in this case the quark matter in the star tends to
become a normal Fermi liquid, for which the description of Iwamoto
which includes Fermi liquid effects, must be adopted.

Fig. \ref{coolsurf} gives the star surface temperature as a function
of time, as obtained by use of Eq. \eqref{surf}. The three curves
refer to the same toy stars as in Fig. \ref{coolcore}. Both diagrams
are obtained for the following values of the parameters: $\mu=500$
MeV, $m_s^2/\mu=140$ MeV, $\Delta_1=0,\, \Delta_2=\Delta_3\simeq 6$
MeV. For unpaired quark matter we use $\alpha_s\simeq 1$, the value
corresponding to $\mu=500$ MeV and $\Lambda_{\rm QCD}=250$ MeV. The
use of perturbative QCD at such small momentum scales is however
questionable. Therefore the results for model II should be
considered with some caution and the curve is plotted only to allow
a comparison with the other models. In particular the  similarity
between the LOFF curve and the unpaired quark curve follows from the
fact that the LOFF phases are gapless, so that the scaling laws
$c_V\sim T$ and $\varepsilon_\nu\sim T^6$ are analogous to those of
the unpaired quark matter. However the cooling curve for unpaired
quark matter depends on the value we assumed for the strong coupling
constant, $\alpha_s\simeq 1$, which is an extrapolation to a regime
where perturbative QCD is less reliable. Therefore the similarity
between the curves of models II and III might be  accidental.

In Fig. \ref{coolcomparison} we compare two models of a star with a
nuclear mantle and a quark core in the LOFF state. The appoximation
used for these LOFF states  were discussed in Sections \ref{LOFF}
and \ref{Sec:1SUmu}. The continuous curve (red online) is for model
III and the dashed black curve is for model IV, i.e. with quarks in
the LOFF2s state. One can note that both curves for the LOFF models
are similar and show a rapid cooling, much faster than for ordinary
stars comprising only nuclear matter. This implies that the results
for models with a LOFF  core are rather robust and, at least for
relatively young stars, the presence of quark matter in the LOFF
state should offer a signature clearly distinct from that of an
ordinary neutron star.

\vskip2cm

\section{Conclusions\label{conclusioni}}
Our study can be summarized as follows. We have computed neutrino
emissivity and specific heat for quark matter in the LOFF state of
QCD in presence of three light quarks. Quark matter has been assumed
to be in weak equilibrium and in a  color and electrically neutral
state. We have considered the simplest ansatz for the condensates
(single plane wave for all LOFF pairings). We have studied two
models of the LOFF state, differing in the approximations used for
the Ginzburg Landau evaluation of the free energy and the gap
equation. Our analysis shows a similarity between the two
approximations, which points to a robustness of our results. We have
used this study for an estimate of the cooling of compact stars with
a nuclear mantle and a quark core.

Which conclusions can we draw from the present study? From Figs.
\ref{coolcore} and \ref{coolsurf} we see that stars with a LOFF core
cool down faster than ordinary neutron stars. This might have
interesting phenomenological consequences. At the present time
observational results on the cooling of pulsars are being
accumulated at an increasing rate (for compilation of data and
comparison between theoretical models and data see, e.g.
\cite{Blaschke:2000dy,Page:2000wt,Prakash:2000jr,Slane:2002ta,kaplan,halpern,Page:2005fq}).
Some data indicate that stars with an age in the range $10^3-10^4$
years have a temperature significantly smaller than what is expected
on the basis of the modified Urca processes. It is difficult however
to infer, from these data, predictions on the star composition, as
the stars may have different masses. As for the impact on our study,
it is useful to repeat here that our analysis should be considered
as preliminary because the identification of the quasiparticle
dispersion laws for the favored crystalline LOFF structures, formed
by more plane waves, is still lacking. This is the reason why
 we have not tried a comparison with observations in this paper.
Nevertheless  some qualitative assessments can be made. Quantum
Chromodynamics predicts that at the densities that can be reached in
the core of compact stars deconfined quark matter should be present,
and, if so, it should be in a color superconductive state, since
Cooper condensation of colored diquarks is energetically favored.
Slow cooling is typical of stars containing only nuclear matter or
of stars with a color superconductive quark core in a gapped phase
(e.g. CFL). If a careful comparison with the data could allow to
rule out slow cooling for star masses in the range we have
considered, this would favor either the presence of condensed mesons
or quark matter in a gapless state in the core, since gapped quarks
emit neutrinos very slowly. Meson condensation also might allow
rapid cooling. However reliable calculations based on the chiral
effective field theory~\cite{Schafer:2005ym} can only be made for
$m_s^2\simeq \mu\Delta$, i.e. far away from the region where the
LOFF state is stable; therefore no direct comparison between the two
phases is available. For intermediate densities the quark normal
state is less favored than color superconductive states for a wide
range of values of the baryonic chemical potential $\mu$ and the
strange quark mass $m_s$. This leaves us with gapless quark phases
and, among them the LOFF state is favored since we know that the
gapless phases with homogeneous gap parameters such as, e.g. the
gCFL or the g2SC phase are instable, while the LOFF phase does not
suffer of a similar instability. Let us finally observe that our
conclusions should remain valid, at least qualitatively, also for
more complex crystalline patterns of the LOFF condensate. Apparently
the fast cooling of relatively young stars with a LOFF quark core is
a consequence of the scaling laws for neutrino emissivity and
specific heat. They depend on the existence of gapless points and
follow from the existence of blocking regions in momentum space.
Since this property is typical of the LOFF state, independently of
detailed form of the condensate, a rapid cooling should be
appropriate not only for the simple ansatz assumed in Eq.
\eqref{cond}, but, more generally, for any LOFF condensate.

\acknowledgements
 We would like to thank M.~Alford, M.~Ciminale,
R.~Gatto,  C.~Kouvaris, A.~Mirizzi, A.~Schmitt, I.~Shovkovy and
Q.~Wang for discussions and comments. One of us (MM) would like to
thank the Barcelona IEEC for the kind hospitality during the
completion of the present work. The work of MM has been supported
by the "Bruno Rossi" fellowship program and by the U.S. Department
of Energy (D.O.E.) under cooperative research agreement
\#DE-FC02-94ER40818.

\appendix
\section{Quasiparticle dispersion laws and mixing coefficients\label{app}}
In this appendix we compute the quasiparticle dispersion laws, the
gappless points and the Bogoliubov coefficients for the electrically
and color neutral LOFF state of QCD with three flavors in beta
equilibrium. In the basis $A=(1,...,9)=
(u_r,\,d_g,\,s_b,\,d_r,\,u_g,\,s_r,\,u_b,\,s_g,\,d_b) $ the gap
matrix is as follows: \be \Delta_{AB}=\left(
\begin{array}{ccccccccc}
  0 & \Delta_3 & \Delta_2 & 0 & 0 & 0 & 0 & 0 & 0 \\
  \Delta_3 & 0 & \Delta_1 & 0 & 0 & 0 & 0 & 0 & 0 \\
  \Delta_2 & \Delta_1 & 0 & 0 & 0 & 0 & 0 & 0 & 0 \\
  0 & 0 & 0 & 0 & -\Delta_3 & 0 & 0 & 0 & 0 \\
  0 & 0 & 0 & -\Delta_3 & 0 & 0 & 0 & 0 & 0 \\
  0 & 0 & 0 & 0 & 0 & 0 & -\Delta_2 & 0 & 0 \\
  0 & 0 & 0 & 0 & 0 & -\Delta_2 & 0 & 0 & 0 \\
  0 & 0 & 0 & 0 & 0 & 0 & 0 & 0 & -\Delta_1 \\
  0 & 0 & 0 & 0 & 0 & 0 & 0 & -\Delta_1 & 0
\end{array}
\right)\, .\label{gapMatrNEW} \ee We consider in detail only the
solution with $\Delta_1=0,~\Delta_2=\Delta_3=\Delta$ (notice that a
factor $e^{2i\bf q_2\cdot r }=e^{2i\bf q_3\cdot r }$ has been
omitted in the entries $\Delta_2,\,\Delta_3$). The case of the
LOFF2s model ($\Delta_1=\Delta_2=0,~\Delta_3\neq 0$), can be treated
in a similar way. Since the matrix \eqref{gapMatrNEW} is
block-diagonal, we consider its various sectors separately.

\subsection{Sector $A=(u_r\,,d_g\,,s_b)$\label{sett123}}
\begin{figure}[ht]
\begin{center}
\includegraphics[width=8cm]{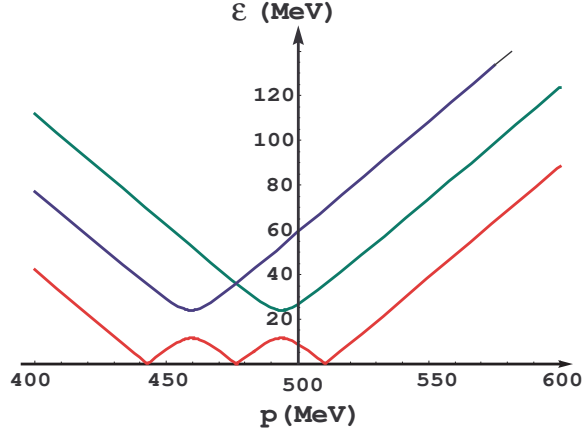}
\end{center}
\caption{\label{fig:emissivita1} (Color online) Dispersion laws for
the quasiparticles given by linear combinations of $u_r\,,d_g\,$ and
$s_b$ quarks. One of the three laws is gapless (red online) with
three gapless momenta. The graph is obtained at $\mu=500$ MeV,
$m_s^2/\mu=140$ MeV and for $\cos\vartheta=0$, where $\cos\vartheta$
is the angle between the quark momentum and the pair momentum 2$\bf
q$. }
\end{figure}

To get the dispersion laws of the quasiparticles one has to find the
poles of the propagator. This corresponds to solve the equation
${\rm det} S^{-1}=0$. The procedure is simplified because one can
write the determinant as a product of various minor determinants. In
particular for  the quasiparticles given by   linear combinations of
$u_r\,,d_g\,$ and $s_b$ quarks one has to evaluate the determinant
of the matrix
 \be\left(\begin{array}{ccc}
    \omega_1 & \Delta & \Delta \\
    \Delta & \tilde\omega_2 & 0 \\
    \Delta & 0 & \tilde\omega_3 \\
  \end{array}
\right)~, \ee where \be \omega_1=\e-p+\mu_1~,~~~~~~ \tilde\omega_2=
\e+p-\mu_2~,~~~~~~~ \tilde\omega_3=\e+p-\mu_3~.\ee The dispersion
laws are obtained by solving the cubic equation in the variable $\e$
\begin{equation}
  (\e-p+\mu_1)(\e+p-\mu_2)(\e+p-\mu_3)-
 \Delta^2[2(\e+p)-\mu_2-\mu_3]~=~0\ .\label{cubic}
\end{equation}
If we define $ y_j=p-\mu_j$ $(j=1,2,3)$ and\bea
b&=&y_1-y_2-y_3~,~~~~~~~~ c=-y_1y_2-y_1y_3+y_2y_3-2\Delta^2~,~~~~~
d=y_1y_2y_3+\Delta^2(y_2+y_3)~,\cr
p&=&c-\frac{b^2}3~,~~~~~~~~~~~~~~~~
q=d-\frac{b\,c}3+\frac{2\,b^3}{27}~,~~~~~~~~~~~~\zeta=
\sqrt[3]{-\,\frac{27\,q}{2}+\frac 3 2\sqrt{12p^3+81q^2}}~,\eea then
the roots of Eq. (\ref{cubic}) are as
follows:\be\epsilon_0=\frac{\zeta}{3w}-\frac {p\, w}\zeta\,+\,\frac
b 3~, ~~~~~\epsilon_a=\,+\,\frac{\zeta\,w}3-\frac
p{\zeta\,w}\,+\frac b 3~,~~~~~
\epsilon_b=-\epsilon_0-\epsilon_a+b~,\label{roots} \ee with
$\displaystyle w=\frac{-1+i\sqrt{3}}{2}$. One can show that only the
quasiparticle of energy $\epsilon_0(p)$ is  gapless, while
$\epsilon_{a,b}$ refer to gapped quasiparticles (this is true for
the reference value we use, $y=140$ MeV, and for any $y<148$ MeV;
for $y>148$ we find that $\epsilon_{a,b}$ are still gapped except
for a tiny region  around the North and the South poles). There can
be one gapless mode ($P_{I})$ or three ($P_{I},P_{II},P_{III}$)
depending on the angle that $2\bf q$ forms with difference of the
pairing quark momenta. The gapless modes are solutions of the
equation\be(P-\mu_1)(P-\mu_2)(P-\mu_3)
+2\Delta^2\left(P-\frac{\mu_2+\mu_3}2\right) =0\ .\label{P123}\ee
The Bogoliubov coefficients given by\bea
B^2_{u_r}(\xi)&=&\frac{(\tilde\omega_2\,\tilde\omega_3)^2}
{f(\tilde\omega_2,\tilde\omega_3,\Delta)}~,\\
B^2_{d_g}(\xi)&=&\frac{(\Delta\,\tilde\omega_3)^2}{f(\tilde\omega_2,\tilde\omega_3,\Delta)}~,\\
B^2_{s_b}(\xi)&=&\frac{(\Delta\,\tilde\omega_2)^2}{f(\tilde\omega_2,\tilde\omega_3,\Delta)}~,
\eea with $ f(\alpha_1,\alpha_2,\alpha_3)=\alpha_1^2\alpha_2^2+
\alpha_1^2\alpha_3^2+ \alpha_2^2\alpha_3^2\ $, represent the
probability that the gapless quasiparticle is respectively  $u_r$,
$d_g$ or $s_b$. We note that the dispersion laws are anisotropic as
they depend on the angle $\vartheta$ that the quark momentum form
with the pair momentum. In Fig. \ref{fig:emissivita1} we plot the
dispersion law for the gapless quasiparticle for $\cos\vartheta=0$.
The quark velocities $v_I$, $v_{II}$ and $v_{III}$ are given by the
slopes of the curve near the gapless points.
\begin{figure}[ht]
\begin{center}
\includegraphics[width=8cm]{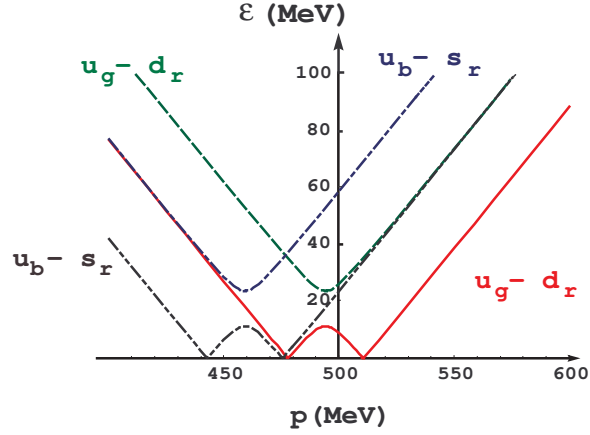}
\end{center}
\caption{\label{fig:emissivita2} (Color online) Solid and double
dotted-dashed lines denote gapless dispersion laws (red and black
online, respectively for $u_g-d_r$ and $u_b-s_r$), dashed and
dot-dashed lines refer to gapped dispersion laws (green and blue
online, respectively for $u_g-d_r$ and $u_b-s_r$). Values of the
parameters as in Fig. \ref{fig:emissivita1}}
\end{figure}
\subsection{Sector $A=(d_r\,,u_g)$\label{sett45}}
In this sector quasiparticles have dispersion laws as follows:\be
 |\,\e^{(d_r\,,u_g)}_\pm| =  \left|
 \pm\delta\m+\sqrt{\left(p-\frac{\mu_4+\mu_5}2\right)^2+\D^2}\right|\,,
\ee where $z=\cos\vartheta$ and $\vartheta$ is the angle
 that the quark momentum form with the pair
momentum. Since \be\delta\mu=\frac{\mu_4-\mu_5}2=\frac y
8\,-\,q\,z\,=\,\frac y 8\left(1\,-\,\frac{z}{z_q}\right)\,~,\ee with
$y$ given in Eq.~\eqref{eq4}. Gapless modes are present for
\be\delta\mu^2~\equiv~\left(\frac y 8\,-\,q\,z\right)^2~>~
\Delta^2~,\label{cond2}\ee therefore the existence of gapless
momenta depends on the value of $z=\cos\vartheta$. The dispersion
law $\e_+$ is gapless for $\delta\mu<0$ whereas $\e_-$ is gapless in
the other case. The corresponding gapless momenta are
\begin{equation}\label{gapless_momenta_45}
  P_\pm^{u_gd_r}=\frac{\mu_4+\mu_5}2\pm\sqrt{\frac{y^2}{64}\left(1-\frac{z}{z_q}\right)^2-\D^2}\,.
\end{equation}
Denoting the gapless dispersion laws with $\epsilon_g$, the mixing
coefficients are given by\be
B_{d_r}~=~\frac{\epsilon_{g}-p+\mu_5}{\sqrt{\Delta^2+(\epsilon_{g}-p+\mu_5)^2}}~,~~~~~
B_{u_g}~=~\frac{\Delta}{\sqrt{\Delta^2+(\epsilon_{g}-p+\mu_5)^2}}~.\ee
Their values at gapless momenta are\bea B^2_{d_r}(P_-^{u_gd_r})&=&
B^2_{u_g}(P_+^{u_gd_r})\,=
\,\frac{1}{2}\left(1-\frac{\sqrt{\delta\mu^2-\Delta^2}}{\delta\mu}\right)\,,
\\
B^2_{d_r}(P_+^{u_gd_r})&=&
B^2_{u_g}(P_-^{u_gd_r})\,=\,\frac{1}{2}\left(1+
\frac{\sqrt{\delta\mu^2-\Delta^2}}{\delta\mu}\right)\ . \eea  We
notice that $\delta\mu>0$ corresponds to $z<z_q$ a condition that is
true almost everywhere (except in a tiny region around the North
pole). 

\subsection{Sector $A=(s_r\,,u_b)$\label{sett67}}
In this sector the quasiparticles have dispersion laws\be
 |\,\e^{(s_r\,,u_b)}_\pm|={\Bigg|}
\delta\mu\pm\sqrt{\left(p-
\frac{\mu_6+\mu_7}2\right)^2+\Delta^2}\Bigg|\ .\label{67}\ee In this
case \be\delta\mu=\frac{\bar\mu_6-\bar\mu_7}2=-\frac{y}{8}~-~q\,z=-
\frac{y}{8}\left(1+\frac{z}{z_q}\right)\ .\ee Gapless modes are
present for \be\delta\mu^2~\equiv~\left(\frac y
8\,+\,q\,z\right)^2~>~ \Delta^2~\label{condbis}\,.\ee The gapless
momenta are
\begin{equation}\label{gapless_momenta_67}
  P_\pm^{u_bs_r}=\frac{\mu_6+\mu_7}2\pm\sqrt{\frac{y^2}{64}\left(1+\frac{z}{z_q}\right)^2-\D^2}\,.
\end{equation}
The values of the Bogoliubov coefficients at the gapless points
are\bea B^2_{s_r}(P_-^{u_bs_r})&=&
B^2_{u_b}(P_+^{u_bs_r})\,=\,\frac{1}{2}\left(1-\frac{\sqrt{\delta\mu^2-\Delta^2}}{\delta\mu}\right)\,,
\\
B^2_{s_r}(P_+^{u_bs_r})&=& B^2_{u_b}(P_-^{u_bs_r})\,=\,\frac{1}{2}
\left(1+\frac{\sqrt{\delta\mu^2-\Delta^2}}{\delta\mu}\right)\ . \eea
Now $\delta\mu>0$ corresponds to $z<-z_q$.
\subsection{Sector $A=(s_g\,,d_b)$\label{sett90}}
There is no mixing in this sector because $\Delta_1=0$, therefore
quasiparticles have dispersion laws
\begin{equation}\label{Disp_law_89}
 |\,\e^{(s_g,\,d_b)}|=|\,p-\m_{8,9}|\,.
\end{equation}
Gapless momenta are
\begin{eqnarray}
 P^{s_g} &=& \m_8=\mu-\frac{5y}{12}\,, \\
  P^{d_b} &=& \m_9=\mu+\frac{y}{12}\,\label{gapless_momenta_bd} \,.
\end{eqnarray}
The mixing coefficients are $ B_{s_g}(P^{s_g})=B_{d_b}(P^{d_b})=1$
and the corresponding velocities are equal to unity.

\end{document}